\documentclass[journal, twoside, web]{ieeecolor}
\usepackage{tikz}
\usepackage{tmi}
\usepackage{cite}
\usepackage{amsmath,amssymb,amsfonts}
\usepackage{algorithmic}
\usepackage{graphicx}
\usepackage{epsfig, epstopdf}
\usepackage{mathtools}
\usepackage{tabularx}
\usepackage{booktabs}
\usepackage{multirow}
\usepackage{makecell}
\usepackage{import}
\usepackage{subfig}
\usepackage{floatrow}

\usepackage{pdfpages}

\usepackage{hyperref}
\usepackage{physics}
\usepackage[all]{hypcap}
\newcommand{\link}[1]{%
    {%
    \def\textcolor##1##2{##2}%
    \def\textbf##1{##1}%
    \def\textit##1{##1}%
    \edef\&{\string&}%
    \edef\_{\string_}%
    \xdef\tmp{\noexpand\href{#1}}}%
 \tmp{#1}%
}

\usepackage{textcomp}
\def\BibTeX{{\rm B\kern-.05em{\sc i\kern-.025em b}\kern-.08em
    T\kern-.1667em\lower.7ex\hbox{E}\kern-.125emX}}
\begin{document}
\title{Efficient Semantic Diffusion Architectures for Model Training on Synthetic Echocardiograms}

\author{
David Stojanovski, Mariana da Silva, Pablo Lamata, Arian Beqiri, Alberto Gomez
\thanks{
Manuscript received X; accepted X . Date of publication X; date of current version X.
This work was supported by the Wellcome/EPSRC Centre for Medical Engineering [WT203148/Z/16/Z] and the National Institute for Health Research (NIHR) Biomedical Research Centre at Guy's and St Thomas' NHS Foundation Trust and King's College London. The views expressed are those of the author(s) and not necessarily those of the NHS, the NIHR or the Department of Health.}
\thanks{ David Stojanovski and Pablo Lamata are with King's College London, London, WC2R 2LS, United Kingdom (e-mail: first.last@kcl.ac.uk).}
\thanks{Arian Beqiri and Alberto Gomez are affiliated with King's College London, and Ultromics Ltd., 4630 Kingsgate, Oxford Business Park South, OX4 2SU, United Kingdom (e-mail: first.last@ultromics.com).}}

\maketitle

\begin{abstract}
We investigate the utility of diffusion generative models to efficiently synthesise datasets that effectively train deep learning models for image analysis. Specifically, we propose novel \(\Gamma\)-distribution Latent Denoising Diffusion Models (LDMs) designed to generate semantically guided synthetic cardiac ultrasound images with improved computational efficiency. We also investigate the potential of using these synthetic images as a replacement for real data in training deep networks for left-ventricular segmentation and binary echocardiogram view classification tasks.

We compared six diffusion models in terms of the computational cost of generating synthetic 2D echo data, the visual realism of the resulting images, and the performance, on real data, of downstream tasks (segmentation and classification) trained using these synthetic echoes. We compare various diffusion strategies and ODE solvers for their impact on segmentation and classification performance. The results show that our proposed architectures significantly reduce computational costs while maintaining or improving downstream task performance compared to state-of-the-art methods. While other diffusion models generated more realistic-looking echo images at higher computational cost, our research suggests that for model training, visual realism is not necessarily related to model performance, and considerable compute costs can be saved by using more efficient models.

\end{abstract}


\section{Introduction}
\label{sec:introduction}
    \IEEEPARstart{E}{chocardiography} (echo) is the most widely used cardiac imaging modality in both clinical and research settings\cite{Dewey2020}, providing real-time information on cardiac structure and function \cite{Soliman-Aboumari2022}. 
    
    Echo acquisition and interpretation is, however, a challenging process that requires extensive expertise and hand-eye coordination to navigate the transducer across the patient's anatomy. This makes resulting images particularly equipment-, patient-, and operator-dependent in terms of view and image quality
    \cite{Beck2024}. This often results in clinically suboptimal images that limit subsequent use \cite{Nicastro2013}, and are liable to high inter- and intra-observer variability in image interpretation, further complicating the extraction of consistent and reliable diagnostic features. These challenges motivate the need for automation and computer-assisted echo image acquisition and analysis.
    
    Deep learning (DL) has shown strong capabilities for automating medical image analysis in various tasks such as classification\cite{Upton2022a, Zhu2021, Xie2020, Sudharson2020}, regression \cite{Zhang2020, Perdios2021, Zhu2021a}, and segmentation \cite{Wei2020, Sfakianakis2023, Oktay2018}. However, they require large, often labelled, datasets. Major challenges in using DL with medical imaging are the availability, access, and privacy preservation of those datasets. This paper investigates the efficient generation of synthetic data as an alternative to real data for training DL models.

\subsection{Related work}
    Advances in generative artificial intelligence have proven the possibility of synthesising new, realistic data. This capability has the potential to improve the robustness and generalisability of medical imaging models \cite{Pinto-Coelho2023}.  By generating diverse synthetic images that include rare cases \cite{Yu2019} or variations in image quality \cite{Escobar2020}, synthetic data can reduce overfitting. While models trained on real data with limited variability often underperform across different patient populations or imaging conditions \cite{Lee2022}, generative models can improve data augmentation strategies (producing more diverse data and reducing the risk of bias) or assist in domain adaptation, helping models trained in data from different hospital equipment \cite{Karani2018} or imaging modalities\cite{Huang2019b} perform well elsewhere.
    
    Generative Adversarial Networks (GANs) and CycleGANs \cite{Jay2017} have until recently been state-of-the-art (SOTA) methods for synthetic image generation \cite{Wolterink2017,Armanious2019}. 
    However, these methods suffer from training instability, vanishing gradients, and mode collapse, limiting their practical utility \cite{Saxena2021}. 
    
    Denoising Diffusion Probabilistic Models (DDPMs) have emerged as a promising alternative, overcoming challenges such as training instability and offering more control over generated outputs \cite{Ho2020}. DDPMs still face limitations; primarily, training and generating datasets using DDPMs has a prohibitive computational expense compared to GANs for many applications and users 
    \cite{Cottier2024}. Recent research has focused on mitigating these costs, for instance by generating images with fewer diffusion steps by more advanced Ordinary Differential Equation (ODE) solvers \cite{Song2021}, architectural innovations to improve convergence rates \cite{Jabri2023}, and knowledge distillation methods to compress large diffusion models into more compact and efficient versions \cite{Karras2024}. The most popular efficiency strategy for SOTA models is to train a Latent Diffusion Model (LDM) \cite{Rombach2022}. LDMs follow a two-stage process in which a Variational Auto-Encoder (VAE) is first trained to map high-dimensional image data into a lower-dimensional, Gaussian-distributed latent space. When the VAE has been trained, it then encodes full-resolution images into a low-resolution latent space on which the diffusion model can be trained. Finally, during inference, the diffusion model denoises an image in the latent space resolution, followed by decoding from the pre-trained VAE model to the full-resolution. 
    
    All of the methods previously discussed have been strictly concerned with maximising image realism, while the utility of synthetic images for AI model development depends on downstream task performance with models trained on the generated images. There is no evidence that realism and utility for training are directly related, and indeed multiple works show that it may not be the case. For example, domain randomisation (a field that seeks to actively generate unrealistic samples \cite{Tobin2017}) is predicated on growing evidence that data augmentation beyond realism can enhance model generalisation \cite{Bengio2011} by effectively creating a training set of synthetic images that act as a superset of real images, and in turn, improve performance in downstream tasks \cite{Billot2023, Tremblay2018}. In this context, we investigate the utility of synthetically generated images by the performance of models trained on them and tested on real data.


    Our research \cite{Stojanovski2023} has demonstrated the effectiveness of DDPMs for the synthesis of semantically guided cardiac ultrasound images. We proposed using a Semantic Diffusion Model (SDM) network, based on the standard DDPM network proposed in \cite{Ho2020}, in conjunction with Spatially Adaptive De-normalisation (SPADE) normalisation blocks in the decoder section of the network for generating cardiac echoes. Using this formulation, synthetic cardiac ultrasound images were generated that were more realistic in appearance than GAN-based methods, while also following the guiding semantic map more closely. However, this method was associated with prohibitively high computational costs compared to GANs and prevented its practical use. The number of steps, \(T\) in the forward process is a crucial hyperparameter in DDPMs. A large \(T\) value ensures that the reverse process closely approximates a Gaussian distribution, making the generative process using Gaussian conditional distributions a good approximation. This rationale supports the use of large \(T\) values, e.g., 1000, as employed by \cite{Ho2020}. However, since each of the \(T\) iterations must be executed sequentially to generate a sample, \(x_0\), sampling from DDPMs is significantly slower compared to other deep generative models. Another issue with using the SPADE blocks in conjunction with LDMs is that there needs to be a clear spatial consistency between the image and associated semantic map, which in the case of Gaussian-distributed VAEs is not necessarily the case.


\subsection{Contributions of this paper}

    In this paper we investigate alternative architectures and solvers to our previously proposed SDM model \cite{Stojanovski2023} that can generate useful images (with respect to training a model for image classification and segmentation) while substantially reducing the computational costs of image generation. In summary, our contributions are:
    
    \begin{itemize}
        \item  Comprehensive comparison of multiple diffusion strategies with SPADE-based semantically guided image synthesis, and the performance of segmentation and classification tasks on real echo images when trained with the synthetically generated data. Included in this, recognising the critical role of ODE solvers in diffusion-based generative models, we integrate and test several ODE solvers to determine their impact on model performance (in both segmentation and classification tasks) and computational efficiency. This assessment aids in identifying optimal solver configurations that balance accuracy and resource utilisation. 
        \item A new \(\Gamma\) Variational Autoencoder (\(\Gamma\)-VAE) tailored for latent diffusion models for echo: a \(\Gamma\) distribution ensures a minimum value of 0, preserving the semantics of the ``background'' (pixels outside the ultrasound sector) and leading to improved latent representation learning and semantic-guided image generation fidelity. The \(\Gamma\) distribution follows the data distribution in the images inside the ultrasound sector more closely. 
    \end{itemize}

\section{Methods}
\label{sec:methods}

    To assess the utility of the synthetically generated images for training DL models, we generate 24 synthetic datasets with six different generative models respectively. Of these six, two are our novel EDM-Lx methods (described below), which are compared to three baseline methods, namely the Elucidating Diffusion Model (EDM) from \cite{Karras2022}, Variance Exploding (VE), and Variance Preserving (VP) from \cite{Song2021}) and to the current SOTA, our previously proposed Semantic Diffusion Model (SDM) \cite{Stojanovski2023}.

\subsection{Overall description of the proposed EDM-Lx model}
\begin{figure}[hbt!]
        \centering
        \centerline{\includegraphics[trim=90 30 0 0, clip, width=\linewidth]{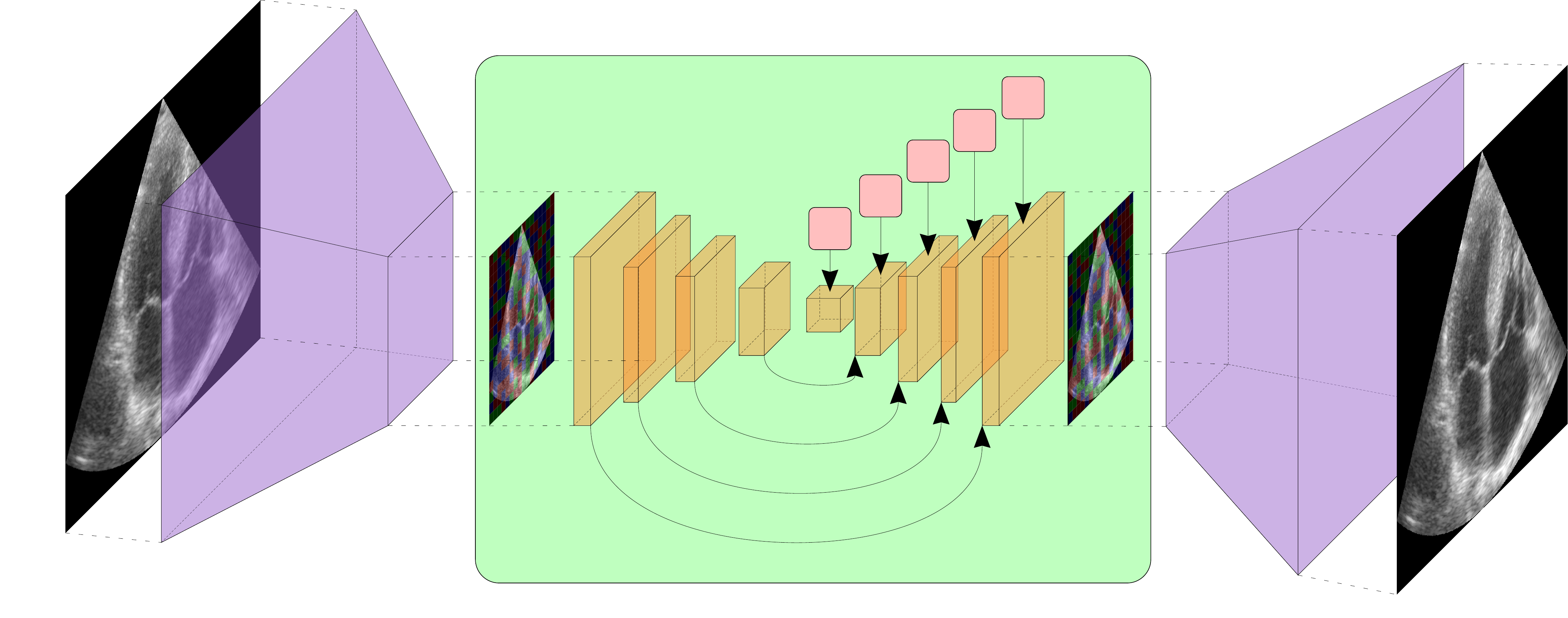}}
        \caption{Architecture diagram of the proposed method. Purple parallelogram: \(\Gamma\)-VAE; green box: diffusion network; red box: SPADE module for semantic conditioning.}
        \label{fig:network arch}
    \end{figure}

    The proposed model is an extension of the EDM model developed by \cite{Karras2024}, where we utilise principles from Latent Diffusion Models (LDMs) and semantically-guided DMs together. We name the model EDM-Lx, where x is the size of the latent space. Figure \ref{fig:network arch} shows a diagram with the proposed architecture. 
    
    The model consists of a new \(\Gamma\)-VAE and a latent diffusion bottleneck. We first pre-train the \(\Gamma\)-VAE for echo image reconstruction and then use the pre-trained \(\Gamma\)-VAE to compress the input image to a smaller resolution for the training of the diffusion model. During image generation, the pre-trained \(\Gamma\)-VAE decoder is used to upscale the output from the diffusion model. Each component is described in turn.

\subsection{\(\Gamma\)-VAE: rationale}
    VAEs are a class of generative models that combine the principles of variational inference and neural networks to learn compact latent representations of data distributions. For the set of all latent variables and parameters, \(\boldsymbol Z\), and the set of all observed variables, \(\boldsymbol X\), a probabilistic model specifies a joint distribution, \(p(\boldsymbol{X}, \boldsymbol{Z})\). The aim is then to find a suitable approximation for the posterior distribution, \(p(\boldsymbol{Z} | \boldsymbol{X})\). The log marginal probability can be decomposed into a summation of the evidence lower bound (ELBO), \(\mathcal{L}(q)\), and Kullback-Leibler divergence, \(D_{\text{KL}}\), as shown in \eqref{eq:log_marginal_prob}. The evidence lower bound can be regarded as the lower bound of the log-likelihood for observed data. The direct calculation of \(p(\boldsymbol{Z} | \boldsymbol{X})\) is computationally expensive and is mostly intractable, so an approximation distribution \(q(\cdot)\) is optimised so that \(q(\boldsymbol{Z} | \boldsymbol{X}) \approx p(\boldsymbol{Z} | \boldsymbol{X})\). Subsequently, it can be found that maximising the lower bound \(\mathcal{L}(q)\) is equivalent to minimising \(D_{\text{KL}}\). Maximising the ELBO would simultaneously aim to obtain an accurate generative and discriminative model \cite{Neal1998}.
    
    \begin{equation}
        \ln p(\boldsymbol{X}) = \mathcal{L}(q) - D_{\text{KL}}(q||p),
        \label{eq:log_marginal_prob}
    \end{equation}
    
    with the following denotations; 
    
    \begin{IEEEeqnarray}{lCr}
        \mathcal{L}(q) &= \int q(\boldsymbol{Z})\ln{\frac{p(\boldsymbol{X}, \boldsymbol{Z})}{q(\boldsymbol{Z})}} d\boldsymbol{Z} \\
        D_{\text{KL}}(q||p) &= \int q(\boldsymbol{Z})\ln{\frac{p(\boldsymbol{Z}| \boldsymbol{X})}{q(\boldsymbol{Z})}} d\boldsymbol{Z}.
    \label{eq:log_marginal}
    \end{IEEEeqnarray}

    As can be seen in \eqref{eq:log_marginal}, the choice for approximating distributions is arbitrary. The most commonly chosen prior is the same as in the original paper proposing VAEs \cite{Kingma2014}, i.e., setting \(p(z)\) to be a normal Gaussian distribution of the form \(X \sim \mathcal{N}(0, 1)\). While this VAE can effectively reconstruct images with good realism, when used in combination with the SPADE diffusion model, the model loses the ability to generate realistic images effectively guided by the semantic maps in the latent space.
    
    We hypothesise that it fails for two reasons. First, the lack of spatial consistency of the latent representation. The guiding semantic map is a multi-label map of cardiac tissues at a lower resolution than the input, and there is nothing in the standard VAE enforcing spatial consistency across resolutions between the latent representation and input image. Note that, while convolutions and pooling operations will propagate local information, this information may be  distributed arbitrarily over many channels of the latent space. 
    
    As a solution for this first reason, we propose that simply restricting the latent space to have one channel prevents spatial features being distributed across channels and that adding a loss term that enforces similarity between the latent representation and a downsampled version of the input encourages consistency between the semantic map and the latent representation. This leaves one remaining inconsistency in the formulation though: the KL divergence promotes that pixels are Gaussian distributed in the latent space, while the input echo image will follow typically a Rayleigh distribution \cite{Deng2011}, resulting in a conflict between the two loss terms. 
    
    
    A second, domain-specific, reason for the failure mode is a particularity of echo images: they display an ultrasound sector region, where image data is defined, and a background region, which is all black and set to a padding value (typically 0). As a result, a desirable property of the latent representation (which should be consistent with the input to be compatible with the guiding semantic map) is that it preserves a fixed background value that ensures proper management of the triangular region. A normal Gaussian distribution enforces a pixel-wise 0-mean, which results in a loss of the background concept after the usual normalisation layers and prior to feeding to the diffusion bottleneck. Conveniently, a \(\Gamma\) distribution is bounded by 0 at its lower end, lending itself as an ideal choice to replace the normal distribution in this case.
    
    Furthermore, substituting the normal Gaussian by a \(\Gamma\) distribution solves the aforementioned conflict between the KL divergence and the latent similarity terms. Even more, as shown next, the KL divergence has a closed-form solution under the \(\Gamma\) distribution. 

\subsection{\(\Gamma\)-VAE: calculation of the KL-divergence}

    
    The PDF for the \(\Gamma\) distribution is:
    
    \begin{align}
        f(x) = \frac{\beta^\alpha}{(\alpha - 1)!}x^{\alpha - 1}e^{-\beta x},&\quad\text{for  } \left[\alpha > 0, \beta > 0\right]
        \label{eq:gamma_pdf}
    \end{align}
    
    where \(\alpha\) and \(\beta\) are the shape and rate parameters for fitting the PDF, and \(x\) is an input value. The KL-divergence between two \(\Gamma\) distributions (disregarding the area outside the ultrasound triangle with the \(\odot \operatorname{Bool}(\text{Sector})\) operation) is: 
    


    \begin{IEEEeqnarray*}{l} 
        D_{\mathrm{KL}}(\alpha_p,\beta_p; \alpha_q, \beta_q)  = (\alpha_p-\alpha_q) \psi(\alpha_p) - \log(\alpha_p-1)! \\ 
        \qquad +\> \log(\alpha_q-1)! + \alpha_q(\log \beta_p - \log \beta_q) + \alpha_p\frac{\beta_q-\beta_p}{\beta_p}, \qquad \\
        \IEEEeqnarraymulticol{1}{l}{
        D_{\mathrm{KL}} = D_{\mathrm{KL}}\odot \operatorname{Bool}(\text{Sector}) } \IEEEyesnumber
        \label{eq:kl_gamma}
    \end{IEEEeqnarray*}

    where \(\alpha_p, \beta_p\) are the prior distribution parameters, \(\alpha_q, \beta_q\) are the approximating distribution parameters, and \(\psi(\cdot)\) is the Digamma function, given by 
    
    \begin{equation}
        \psi(x) = \frac{d}{dx} \ln(x-1)!
    \end{equation}
    
    During the training phase of the \(\Gamma\)-VAE, the reparametrisation trick is used\cite{Kingma2014}. However, during the diffusion model training phase, we do not sample from the \(\Gamma\) distribution; instead, the latent representation corresponding to the expected value of the distribution is used. The expectation for a random sample \(X \sim \Gamma(\alpha, \beta)\) is given by
    
    \begin{equation}
        \mathbb{E}(X) = \frac{\alpha}{\beta}.
    \end{equation}

    This ensures each segmentation mask provided to the diffusion model has a unique corresponding latent representation. 
    
    The final \(\Gamma\)-VAE loss function for an input image, \(X\) is:
    
    \begin{align}
        \begin{split}
            \mathcal{L} &= \lambda_1 \| X - Y \|^2_2 + \lambda_2 \mathcal{P}(X,Y) \\
            &+ \lambda_3 D_{KL} + \lambda_4 \| x - y \|^2_2,
        \end{split}
    \end{align}

    where \(\lambda_{(\cdot)}\) are loss weighting factors, \(Y\), is the predicted image reconstruction, \(\mathcal{P}(\cdot)\), is the perceptual loss function \cite{Zhang2018}, \(x\) and \(y\) are the subsampled resolution image and the latent prediction image, respectively.

\subsection{Diffusion model: rationale}
    In latent diffusion models (LDMs), the latent space of the VAE is replaced by a diffusion process. A diffusion process can be understood as a Markov chain of \(T\) steps. In the forward process, also known as the diffusion process, small amounts of noise are iteratively added to the data. This process continues until the data is transformed into pure noise. Formally, for a given distribution \(x_0 \sim q(x_0)\), the forward noising process, \(q\), can be defined as:
    
    \begin{equation}
        q(x_t | x_{t-1}) \coloneq \mathcal{N}(x_t; \sqrt{1 - \beta_t} x_{t-1}, \beta_t \textbf{I}).
    \end{equation}
    
    where \(x_t\) represents the data at timestep \(t\), \(\beta_t\) is a variance schedule that controls the amount of noise added at each step, and \(\mathcal{N(\cdot)}\) denotes a Gaussian distribution. The reverse distribution, \(q(x_{t-1}|x_t)\), depends upon the entire data distribution and so is approximated in a network using:
    
    \begin{equation}
          p_{\theta}(x_{t-1} \mid x_{t}) := \mathcal{N}(x_{t-1}; \mu_{\theta}(x_{t}, t), \Sigma_{\theta}(x_{t}, t)),
        \label{eq:placeholder_label}
    \end{equation}
    where \(\theta\) represents the learnable parameters of the model.

    The probability flow ordinary differential equation (ODE) is a continuous-time formulation of the diffusion process. This ODE continuously adjusts the noise level of the input when moving forward (increasing noise) or backward (decreasing noise) in time. 
    To fully specify the probability flow ODE, a noise schedule function, \(\sigma(t)\), that defines the desired noise level at each time point, \(t\), must be chosen. This noise schedule determines how the noise is gradually added or removed from the image during the diffusion process. The differential step in the diffusion process can be given by
    
    \begin{equation}
        d\boldsymbol{x} = -\dot{\sigma}(t) \sigma(t) \nabla_x \log p(\boldsymbol{x}; \sigma(t)) \, dt,
        \label{eq:diff_step}
    \end{equation}
    
    where \(\log p(\boldsymbol{x}; \sigma)\) is defined as the score function \cite{Hyvarinen2009}, a vector gradient towards a higher density of data for a given noise level. The diffusion network can be regarded as a denoising function, \(\mathcal{D}(\boldsymbol{x}; \sigma)\), which minimises the expected \(L_2\) norm denoising error, i.e.
    
    \begin{equation}
            \min \| \mathcal{D}(\boldsymbol{x+n}; \sigma) - \boldsymbol{x} \|^2_2
    \end{equation}
    
    where \(\boldsymbol{x}\) is the input image and \(\textbf{n}\) is the added noise. The alternative definition of the score function can be defined as 
    
    \begin{equation}
            \nabla_{\boldsymbol{x}} \log p(\boldsymbol{x}; \sigma) = \frac{(\mathcal{D}(\boldsymbol{x}; \sigma) - \boldsymbol{x})}{\sigma^2}.
            \label{eq:alt_score_func}
    \end{equation}

    We are strictly concerned with conditional image generation where our conditioning is the corresponding semantic label maps, \(\Phi\). Using this fact and substituting \eqref{eq:alt_score_func} into \eqref{eq:diff_step} we get the general form of the ODE equation that can be solved with any desired ODE solver:
    \begin{equation}
        d\boldsymbol{x} = -\dot{\sigma}(t) \sigma(t) \frac{(\mathcal{D}(\boldsymbol{x|\Phi}; \sigma) - \boldsymbol{x})}{\sigma^2} dt.
    \end{equation}
    
    Numerical solutions of ODEs involve approximating the true solution path. Each approximation step introduces a truncation error, which accumulates over multiple steps. The error typically increases faster than linearly with step size, so increasing the number of steps improves accuracy. Euler's method, a commonly used solver, produces \(\mathcal{O}\left(h^2\right)\) local error with respect to step size \(h\). Higher-order Runge-Kutta methods offer better error scaling but require more evaluations of the diffusion model per step. Heun's method incorporates an additional correction step between steps at \(t_i\) and \(t_{i+1}\) \cite{Suli2003}. This correction yields \(\mathcal{O}\left(h^3\right)\) local error at the expense of one additional Neural Function Evaluation (NFE) per step. We thus follow the recommendations in \cite{Karras2022} and use the Heun method.

\subsection{Diffusion model: semantic conditioning}
    The Spatially Adaptive De-normalization (SPADE) method was proposed to provide spatially-variant normalisation, designed to adapt to the semantic content of input images \cite{Park2019}. The SPADE module for a layer of a convolutional neural network is formally defined as
    
    \begin{equation}
        \gamma_{c,y,x}(\mathbf{m}) \cdot \frac{h_{n,c,y,x} - \mu_c}{\sigma_c} + \beta_{c,y,x}(\mathbf{m}),
        \label{eq:spade}
    \end{equation}
    
    where \(\mathbf{m}\) is the given segmentation mask, \(\gamma(\cdot)\) and \(\beta(\cdot)\) are the learned modulation parameters of the network layer in the given channel \(c\), with spatial position \(x,y\) ; \(h\) is the activation at the site before normalisation for batch size \(n\), and \(\mu_c\) and \(\sigma_c\) are the mean and standard deviation of the activations in the channel.
    
    Consequently, by substituting all commonplace normalisation layers with SPADE blocks, it becomes possible to effectively propagate semantic information through the network, which was shown to improve visual fidelity and spatial congruence \cite{Park2019}.

\section{Materials and experiments}
\label{sec:experiments}

\subsection{Collection and preparation of real data}
    We utilised the CAMUS echocardiography dataset \cite{Leclerc2019}, comprising 500 patients with semantic (segmentation) maps with three labels: left ventricle (LV) myocardium, endocardium, and left atrium (LA) area in apical four-chamber (A4C) and two-chamber (A2C) views at end-diastole (ED) and end-systole (ES). This dataset consists of a mixture of healthy and diseased patients, acquired across multiple sites in France. After reserving 50 patients for testing purposes on the downstream tasks (described later), we used the remaining 450 for training and validating the generative models. We added a fourth label to the semantic maps with the ultrasound sector (triangle) through simple thresholding on the original images. The native resolution of that dataset is $256 \times 256$ pixels.
    
\subsection{Architectural details for the generative models}

\subsubsection{Diffusion models}
    The EDM, VE, and VP configurations were all trained and tested at the full \(256 \times 256\) resolution. Subsequently, the EDM model was also trained in the \(\Gamma\)-VAE latent configuration at latent resolutions of \(64 \times 64\) and \(128 \times 128\) for the EDM-L64 and EDM-L128 models, respectively. 
    

\subsubsection{Variational Autoencoders}
    The EDM model with $\Gamma$-VAEs encoded inputs into smaller representations of \(128\times128\) and \(64\times64\) pixels with 2 and 3 encoder layers respectively. The encoder and decoder were symmetric in structure with \( N_{params} = layer \cdot 2^{6}\). In practical terms, appropriate values of \(\alpha\) and \(\beta\) for our \(\Gamma\) distribution parameters are needed; for the CAMUS training set, we found, via grid search, appropriate values to be \(3.75\) and \(10.8\) respectively.

\subsection{Downstream tasks for data evaluation}
    Downstream tasks were trained using synthetic data (using the same seantic labels for all generative models). Two downstream tasks were used to assess the quality of the synthetic data: a segmentation task and a view classification task. Both models were trained using a $256 \times 256$ size input, for 250 epochs using the Adam optimizer\cite{Kingma2015b} with a learning rate of $1 \times 10^{-4}$.
    
\subsubsection{Segmentation Model}
    The segmentation model used was the MONAI \cite{Pinaya2023} implementation of nnU-Net proposed by Hatamizadeh et al. \cite{Isensee2021}. This network has been shown to have excellent performance on the CAMUS dataset and can act as a common benchmark.

\subsubsection{Classification Model}
    We chose the EfficientNetV2-Small \cite{Tan2021} model to implement view classification (A2C vs A4C). This model is publicly available via PyTorch and provides a standardized baseline that can be easily reproduced across different research environments. 

\subsection{Experiments}

We first provide a brief insight into the appearance of the syntheticly generated images and specifically illustrate what the images generated with the different models look like for increasing number of denoising iterations. 

Then we carry out a quantitative analysis of the utility of the synthetic images when used for training the downstream task models. To that end, we first generate five new synthetic datasets, of 9000 images each, using the five generative models (VE, VP, EDM, EDM-L128 and EDM-L64). All five models took the same inputs: 9000 semantic maps. These maps come from the 400 + 50 CAMUS patients used for training and validating the generative models, and adding 5 random transformations to each as described in \cite{Stojanovski2023}. Note this approach left out 50 patients from the CAMUS dataset for testing the downstream task models.

We trained the downstream task models (segmentation and classification) using the different synthetic datasets until convergence. For the segmentation task, we assessed the segmentation task with the Dice coefficient \cite{Dice1945} and the mean Hausdorff distance (HD) \cite{Huttenlocher1992} of the left ventricular endocardium. These are defined in \eqref{eq:dice} and \eqref{eq:hd_metric}, respectively.
\begin{equation}
    \text{Dice} = \frac{2 |X \cap Y|}{|X| + |Y|}
    \label{eq:dice}
\end{equation}
\begin{align}
    \begin{split}
        H(X, Y) &= \max(h(X, Y), h(Y, X)) \\
        h(X, Y) &= \max_{x \in X} \min_{y \in Y} \|x - y\|
    \label{eq:hd_metric}
    \end{split}
\end{align}
where \(X\) and \(Y\) represent the pixels with non-zero values in the prediction and ground truth segmentation, respectively.

For the classification task, we used the classification accuracy, defined for the binary classification task as
\begin{equation}
    \text{A} = \frac{TP + TN}{TP + TN + FP + FN}.
\end{equation}
where TP, TN, FP, and FN are true positive, true negative, false positive, and false negative classifications, respectively.

Both the classification and segmentation experiments were bootstrapped by taking a random \(80\%\) subset of the test set and repeating this for 1000 iterations. The values reported are the mean values and standard deviations over 1000 iterations.

Last, we carried out a computational cost analysis by measuring the throughput for each generative model (in images per second) on the same compute platform and correlated that with the number of Neural Function Evaluations (NFEs) and downstream task performance.


\subsection{Further implementation details and code availability}
All models were trained on 6$\times$Nvidia 40GB A100 graphics processing units using Pytorch 2.4 \cite{Paszke2019a}. The diffusion models were trained for $5\times10^{5}$ images with global batch sizes of 24, 54, and 120, for the $256\times256$ native resolution, $128\times128$, and $64\times64$ latent resolution models, respectively. The \(\Gamma\)-VAE models were trained for 200 epochs and a batch size of 12. The model throughput was calculated on an Nvidia RTX 3090 GPU. Code and data is available at \link{https://github.com/david-stojanovski/EDMLX}.

During training for the diffusion models, only geometric transformations were applied. No colorimetric transforms were performed, in order not to encourage the generation of images that are outside the realistic bounds of the physics constraints of ultrasound imaging. The autoencoder and segmentation network training included both colourimetric and geometric transforms. Moreover, the loss calculation for the generative models was restricted to the sector region of the ultrasound image. For each loss term, we first calculate the loss on a per-pixel basis, but before the sum reduction, we perform an elementwise multiplication between the per-pixel loss and a binarised guiding segmentation mask. To turn the sum into a mean (which we found more numerically stable for training), we divided by the number of in-sector pixels. 

\section{Results}
\label{sec:results}


\subsection{Qualitative Results: Realism}

\begin{figure*}[hp!]
    \centerline{\includegraphics[width=0.95\textwidth]{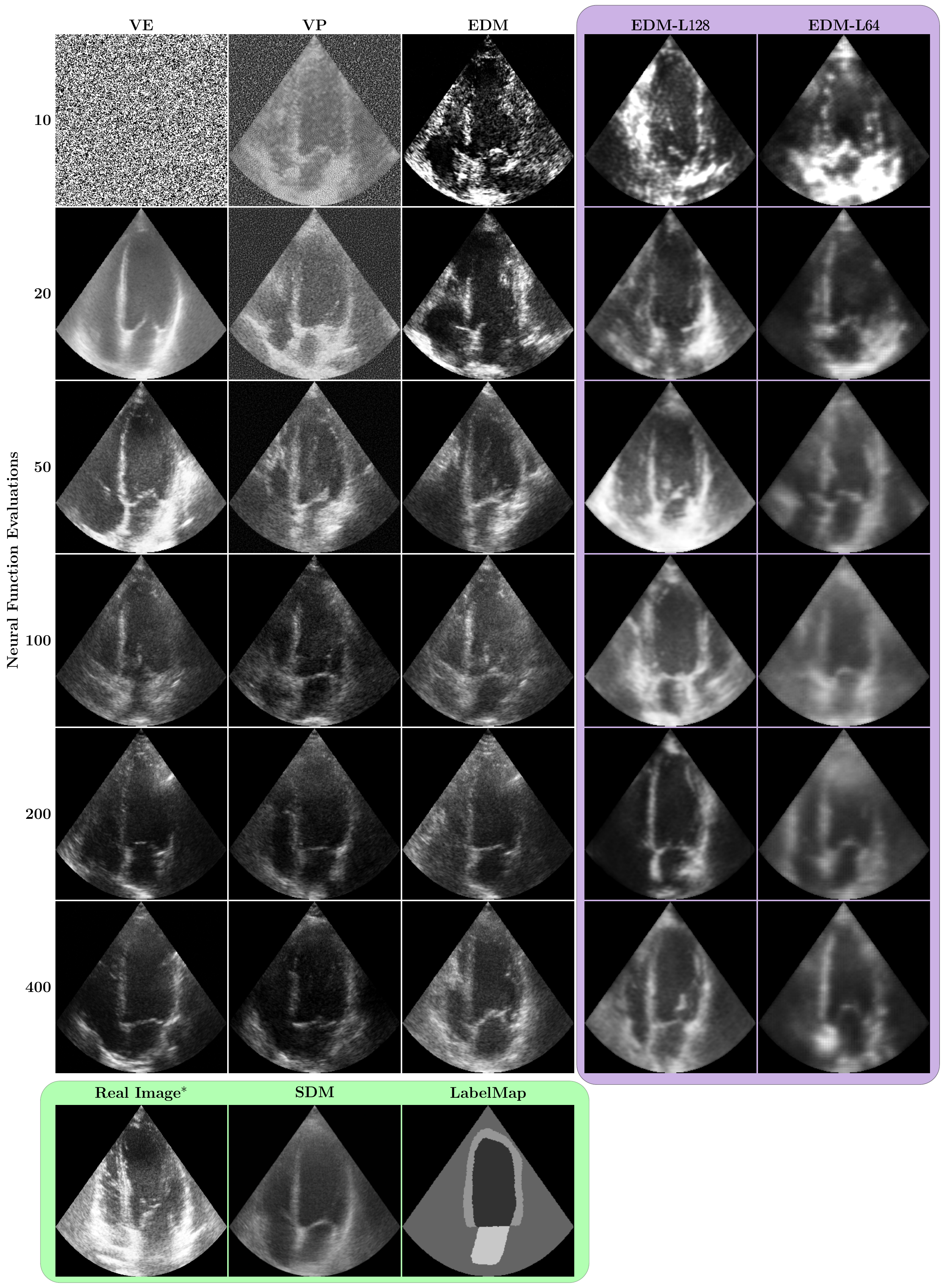}}
    \caption{Example of 4 chamber generated images from all models across all timesteps. Each image was generated using the same guiding segmentation mask as input (shown in green box). Real Image\(^*\): Image from which the labelmap was derived. Purple Box: Latent Models.}
    \label{fig:all_model_examples}
\end{figure*}

Fig. \ref{fig:all_model_examples} shows an example A4C ED image generated by all models across an increasing number of neural function evaluations. We found that this example is representative of the entire dataset in terms of appearance and variation across models.

\newpage
Visually, image realism, considering the presence of correct anatomical structures with a sharp appearance and contrast, generally peaks at 50 NFEs across all models. All full-resolution models (i.e., VE, VP, and EDM) can be seen to generate highly realistic examples of ultrasound images after 50 NFEs, albeit with varying styles. The latent diffusion models introduce blurriness, a characteristic expected from the VAE reconstruction that is more intense the lower the resolution of the latent space. 

The appearance of the VE model at 20 NFEs has a strong visual resemblance to the SDM models from our previous work's \cite{Stojanovski2023} generated images at 1000 NFEs. 

\subsection{Quantitative Results: Downstream Tasks}
\subsubsection{Performance on real data of the synthetic-trained segmenter}\label{sec:dice_score}

\begin{figure}[htb!]
\centering
\sidesubfloat[]{\label{fig:mean_dice_vs_nfe}\includegraphics[width=\columnwidth]{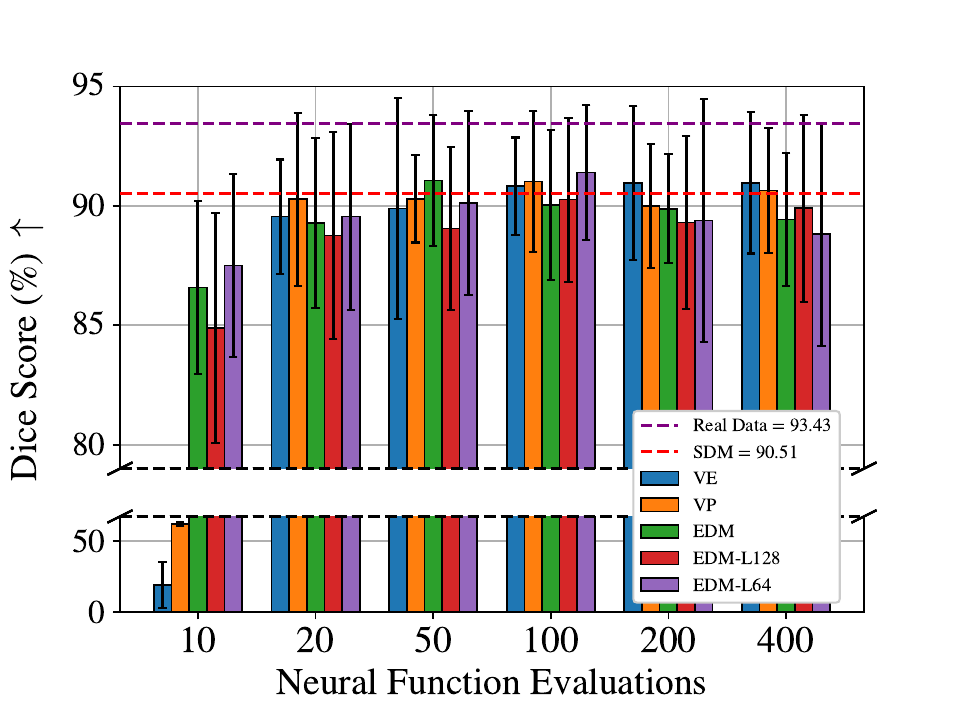}} \\
\sidesubfloat[]{\label{fig:mean_hd_vs_nfe}\includegraphics[width=\columnwidth]{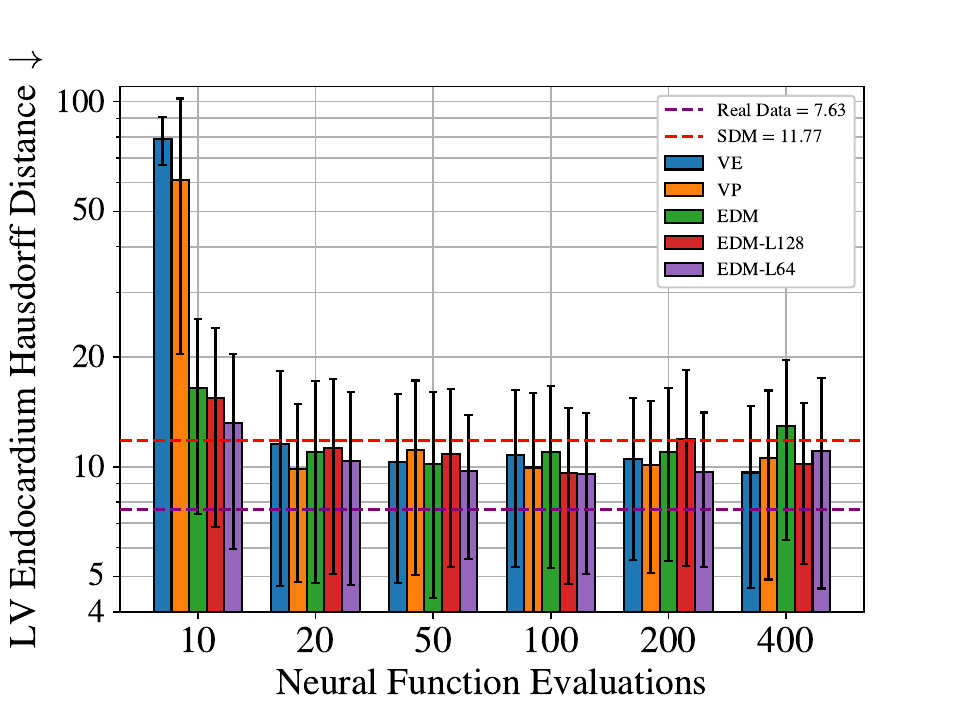}} \\
\sidesubfloat[]{\label{fig:mean_acc_vs_nfe}\includegraphics[width=\columnwidth]{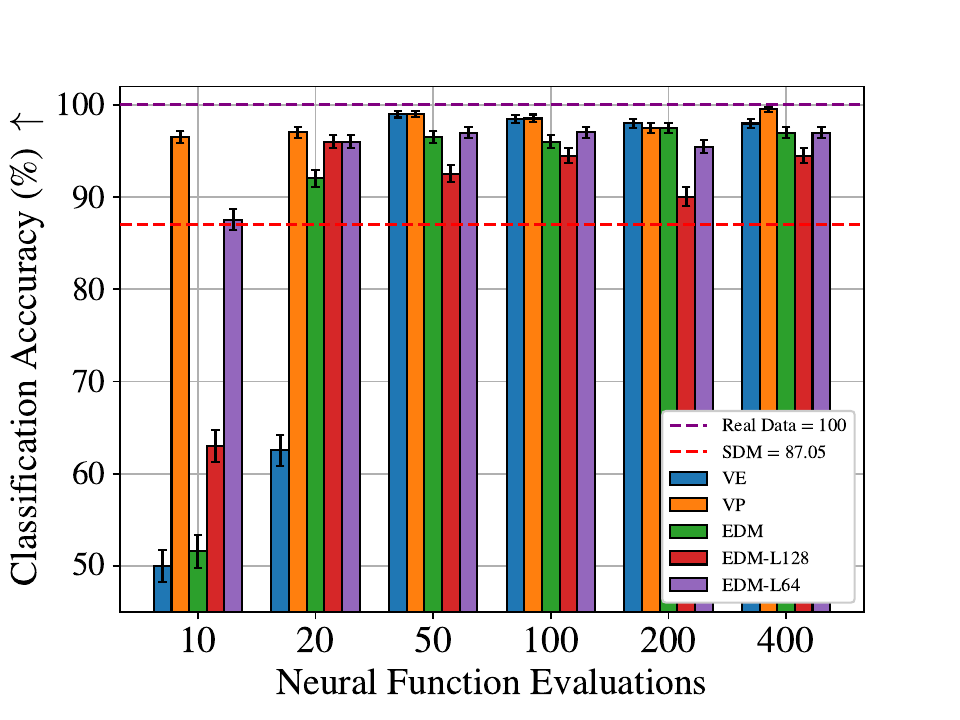}}
\caption{Comparing model performance on segmentation tasks and classification tasks.}
\end{figure}

Fig. \ref{fig:mean_dice_vs_nfe} shows the mean Dice score achieved by the diffusion models against the number of NFEs. The red dashed line indicates the best possible performance when the model is trained on real data, and the purple dashed line indicates the performance achieved when training on synthetic images generated by the current SOTA (our previously proposed SDM method from \cite{Stojanovski2023}). 
 All EDM-based models achieved Dice $>88$ even at 20 NFEs, with all models converging from this point and displaying no significant differences. 

\floatsetup[figure]{subcapbesideposition=top}

Fig. \ref{fig:mean_hd_vs_nfe} shows the Hausdorff distance of all models for the LV endocardium. 
Though the SOTA SDM model displays an equivalent performance for the Dice score, it is surpassed by all models from 20 NFEs and onwards when evaluating the Hausdorff distance for the LV Endocardium. The HD measures the furthest point of dissimilarity and is sensitive to outliers or localised errors. Even a small portion of one boundary deviating significantly from the other can result in a large HD. The lower HD of our models shows that there is greater consistency in the structure of the LV endocardium within the generated images. The EDM-L64 model is consistently among the lowest Hausdorff distances measured on the test set. 

\subsubsection{Performance on real data of the synthetic-trained classifier}
\label{sec:hausdorff_dist}
Fig. \ref{fig:mean_acc_vs_nfe} shows the mean accuracy for performing view classification between A2C or A4C. 
With 10 or more NFEs, the VP and EDM-L64 models can supersede the performance of the previous SOTA SDM model, and by 50 NFEs, all models have outperformed SDM. The performance difference between real data and synthetic data is marginal once models have converged. These results show that the visual features that a network uses for segmentation and classification are inherently different. 

\subsection{Quantitative Results: Computational Efficiency}
\label{sec:bootstrap_acc}

\begin{figure}[htb!]
\centering
    \centerline{\includegraphics[width=\linewidth]{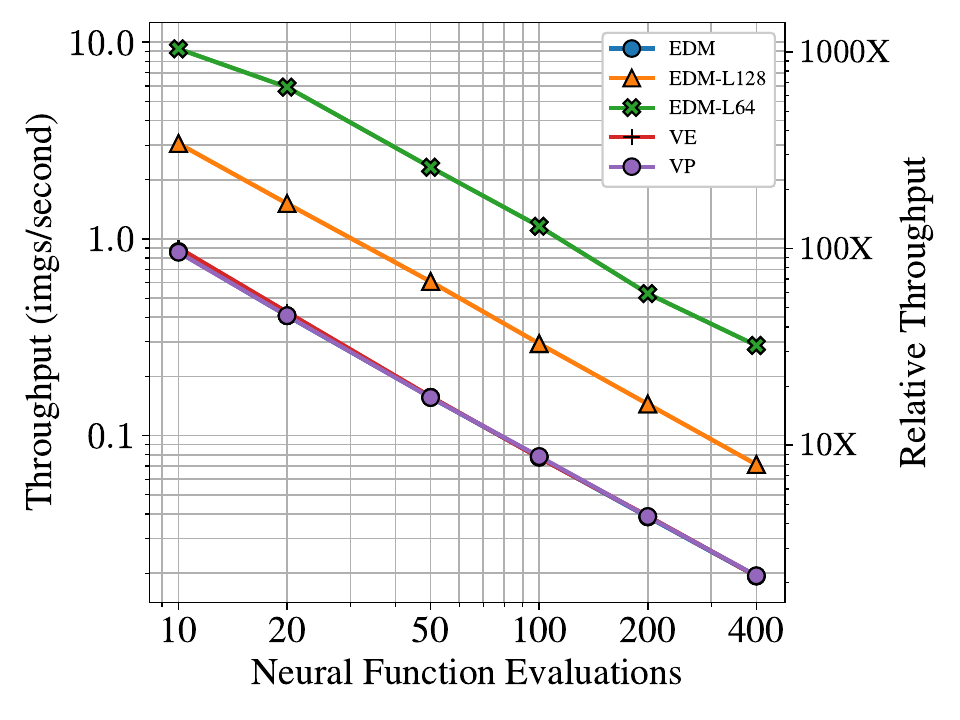}}
    \caption{Throughput of models and relative throughput when compared against SDM. Note: EDM, VE, and VP are functionally identical in performance.}
    \label{fig:nfe_vs_throughput}
\end{figure}

Figure \ref{fig:nfe_vs_throughput} shows the throughput (both in absolute terms in the left y-axis and in relative terms to the SDM method, on the right y-axis) of the 5 competing generative models. Note that the EDM, VE, and VP methods practically overlap, showing effectively an indistinguishable computational efficiency. The throughput decreases logarithmically with the number of NFEs in all cases. Independently of the number of NFEs, the throughput of the EDM-L64 model (the fastest one) is \(10\text{X}\) that of VP, VE, or EDM and \(3\text{X}\) that of EDM-L128.

The peak Dice score obtained was from the EDM-L64 dataset at 100 NFEs. This is the most efficient general model and can perform better than the previous state-of-the-art SDM with more than a \(100\text{X}\) decrease in image generation compute. For classification task purposes, the reduction in computational power required is even greater as superior performance can be obtained with the EDM-L64 at 10 NFEs, which is a greater than \(1000\text{X}\) decrease in relation to SDM. 
Focusing on 100 NFEs, where EDM-L64 achieves top performance in segmentation and classification, to match the throughput, EDM-L128 can run up to around 25 NFEs (where, in terms of performance, it is close but not yet converged to peak), and the other models would not be able to run even 10 NFEs where performance is significantly below peak.


\section{Discussion}

A much more efficient (up to 1,000 times with respect to the SOTA method with 1000 NFEs) generative model is presented. While arguably the blurriness introduced by VAE-based methods can be seen as a lower visually perceived realism, quantitative results show that there is no significant drop in downstream task performance (i.e., image quality for image segmentation and view classification). 

The highest Dice score is obtained with EDM-L64 (the model with the lowest resolution latent space) at 100 NFEs, suggesting that visually perceived image realism and model-synthesised image quality are not necessarily the same. Further, the results in Sections \ref{sec:dice_score}, \ref{sec:hausdorff_dist}, \ref{sec:bootstrap_acc} suggest that there may not be a correlation between image realism and downstream task performance for either segmentation or classification. This supports the idea of domain randomisation introduced earlier and the need for establishing image quality metrics that reflect the quality of synthetic data for a specific intended use. This is a considerable challenge since human-perceived realism features will not necessarily be good image quality metrics. 

We observe that both Dice and Haussdorf distance graphs have a subtle trend to peak at around 100 NFEs and then show a mild regression in performance towards higher NFEs. While we could not establish the reason for this, and visually there is little or no difference in the generated images beyond 50 NFEs, we hypothesise that the noise present at 50-100 NFEs acts as a regulariser that allows for greater generalizability, and this might be lost with more NFEs. Proving this hypothesis is out of the scope of this paper.

In both segmentation and classification tasks, it is seen that at low NFEs (up to 20), the EDM model and variants follow the trend that the lower the resolution, the better the utility to train downstream tasks. While the performance achieved at this relatively low number of NFEs is far from the peak performance, this raises the question, to be addressed in future work, of whether a larger number of synthetic images trained at lower NFEs may outperform the proposed setups.

Our results suggest that LDMs can produce useful synthetic images that are as useful as images generated by full-resolution models at a computational cost of 10 times lower (in the case of EDM-L64). This translates directly into financial savings when using, for example, cloud computing services which charge per time of use.


\section{Conclusion}
The highly effective combination of SPADE blocks and diffusion models can be applied to different architectures and ODE solver sampling schemes, as well as to latent diffusion models. Such models generate images that, while visually appearing of variable realism, are efficient for training models for downstream tasks that then perform well on real data and, when using LDMs, achieve large computational cost savings.

\bibliography{my_bib}

\begin{thebibliography}{10}

\bibitem{Dewey2020}
M.~Dewey, M.~Siebes, M.~Kachelrie{\ss}, K.~F. Kofoed, P.~Maurovich-Horvat, K.~Nikolaou, W.~Bai, A.~Kofler, R.~Manka, S.~Kozerke, A.~Chiribiri, T.~Schaeffter, F.~Michallek, F.~Bengel, S.~Nekolla, P.~Knaapen, M.~Lubberink, R.~Senior, M.~X. Tang, J.~J. Piek, T.~van~de Hoef, J.~Martens, and L.~Schreiber, ``{Clinical quantitative cardiac imaging for the assessment of myocardial ischaemia},'' {\em Nature Reviews Cardiology}, vol.~17, no.~7, pp.~427--450, 2020.

\bibitem{Soliman-Aboumari2022}
H.~Soliman-Aboumari, S.~S. Joshi, M.~Cameli, B.~Michalski, R.~Manka, K.~Haugaa, A.~Demirkiran, T.~Podlesnikar, R.~Jurcut, D.~Muraru, L.~P. Badano, and M.~R. Dweck, ``{EACVI survey on the multi-modality imaging assessment of the right heart},'' {\em European Heart Journal Cardiovascular Imaging}, vol.~23, pp.~1417--1422, nov 2022.

\bibitem{Beck2024}
S.~Beck, H.~{Aziz Shamri}, S.~Coffey, M.~Anakin, and G.~Whalley, ``{Image quality and technical limitations in emergency department cardiac point-of-care ultrasound: A retrospective cohort study},'' {\em EMA - Emergency Medicine Australasia}, vol.~36, no.~2, pp.~295--301, 2024.

\bibitem{Nicastro2013}
I.~Nicastro, V.~Barletta, L.~Conte, I.~Fabiani, A.~Morgantini, G.~Lastrucci, and V.~Bello, ``{Professional education, training and role of the cardiac sonographer in different countries},'' {\em Journal of Cardiovascular Echography}, vol.~23, no.~1, pp.~18--23, 2013.

\bibitem{Upton2022a}
R.~Upton, A.~Mumith, A.~Beqiri, A.~Parker, W.~Hawkes, S.~Gao, M.~Porumb, R.~Sarwar, P.~Marques, D.~Markham, J.~Kenworthy, J.~M. O'Driscoll, N.~Hassanali, K.~Groves, C.~Dockerill, W.~Woodward, M.~Alsharqi, A.~McCourt, E.~H. Wilkes, S.~B. Heitner, M.~Yadava, D.~Stojanovski, P.~Lamata, G.~Woodward, and P.~Leeson, ``{Automated Echocardiographic Detection of Severe Coronary Artery Disease Using Artificial Intelligence},'' {\em JACC: Cardiovascular Imaging}, vol.~15, no.~5, pp.~715--727, 2022.

\bibitem{Zhu2021}
Y.-C. Zhu, P.-F. Jin, J.~Bao, Q.~Jiang, and X.~Wang, ``{Thyroid ultrasound image classification using a convolutional neural network},'' {\em Annals of Translational Medicine}, vol.~9, no.~20, pp.~1526--1526, 2021.

\bibitem{Xie2020}
J.~Xie, X.~Song, W.~Zhang, Q.~Dong, Y.~Wang, F.~Li, and C.~Wan, ``{A novel approach with dual-sampling convolutional neural network for ultrasound image classification of breast tumors},'' {\em Physics in Medicine and Biology}, vol.~65, no.~24, 2020.

\bibitem{Sudharson2020}
S.~Sudharson and P.~Kokil, ``{An ensemble of deep neural networks for kidney ultrasound image classification},'' {\em Computer Methods and Programs in Biomedicine}, vol.~197, no.~2020, 2020.

\bibitem{Zhang2020}
J.~Zhang, C.~Petitjean, P.~Lopez, and S.~Ainouz, ``{Direct estimation of fetal head circumference from ultrasound images based on regression CNN},'' {\em Proceedings of Machine Learning Research}, vol.~121, pp.~914--922, 2020.

\bibitem{Perdios2021}
D.~Perdios, M.~Vonlanthen, F.~Martinez, M.~Arditi, and J.~P. Thiran, ``{CNN-Based Ultrasound Image Reconstruction for Ultrafast Displacement Tracking},'' {\em IEEE Transactions on Medical Imaging}, vol.~40, no.~3, pp.~1078--1089, 2021.

\bibitem{Zhu2021a}
F.~Zhu, M.~Liu, F.~Wang, D.~Qiu, R.~Li, and C.~Dai, ``{Automatic measurement of fetal femur length in ultrasound images: A comparison of random forest regression model and SegNet},'' {\em Mathematical Biosciences and Engineering}, vol.~18, no.~6, pp.~7790--7805, 2021.

\bibitem{Wei2020}
H.~Wei, H.~Cao, Y.~Cao, Y.~Zhou, W.~Xue, D.~Ni, and S.~Li, ``{Temporal-Consistent Segmentation of Echocardiography with Co-learning from Appearance and Shape},'' in {\em Lecture Notes in Computer Science (including subseries Lecture Notes in Artificial Intelligence and Lecture Notes in Bioinformatics)}, vol.~12262 LNCS, pp.~623--632, 2020.

\bibitem{Sfakianakis2023}
C.~Sfakianakis, G.~Simantiris, and G.~Tziritas, ``{GUDU: Geometrically-constrained Ultrasound Data augmentation in U-Net for echocardiography semantic segmentation},'' {\em Biomedical Signal Processing and Control}, vol.~82, no.~August 2022, p.~104557, 2023.

\bibitem{Oktay2018}
O.~Oktay, E.~Ferrante, K.~Kamnitsas, M.~Heinrich, W.~Bai, J.~Caballero, S.~A. Cook, A.~{De Marvao}, T.~Dawes, D.~P. O'Regan, B.~Kainz, B.~Glocker, and D.~Rueckert, ``{Anatomically Constrained Neural Networks (ACNNs): Application to Cardiac Image Enhancement and Segmentation},'' {\em IEEE Transactions on Medical Imaging}, vol.~37, no.~2, pp.~384--395, 2018.

\bibitem{Pinto-Coelho2023}
L.~Pinto-Coelho, ``{How Artificial Intelligence Is Shaping Medical Imaging Technology: A Survey of Innovations and Applications},'' {\em Bioengineering}, vol.~10, no.~12, 2023.

\bibitem{Yu2019}
K.~Yu, Y.~Wang, Y.~Cai, C.~Xiao, E.~Zhao, L.~Glass, and J.~Sun, ``{Rare Disease Detection by Sequence Modeling with Generative Adversarial Networks},'' 2019.

\bibitem{Escobar2020}
M.~Escobar, A.~Castillo, A.~Romero, and P.~Arbel{\'{a}}ez, ``{Ultragan: Ultrasound enhancement through adversarial generation},'' {\em Lecture Notes in Computer Science (including subseries Lecture Notes in Artificial Intelligence and Lecture Notes in Bioinformatics)}, vol.~12417 LNCS, pp.~120--130, 2020.

\bibitem{Lee2022}
T.~Lee, E.~Puyol-Ant{\'{o}}n, B.~Ruijsink, M.~Shi, and A.~P. King, ``{A Systematic Study of Race and Sex Bias in CNN-Based Cardiac MR Segmentation},'' {\em Lecture Notes in Computer Science (including subseries Lecture Notes in Artificial Intelligence and Lecture Notes in Bioinformatics)}, vol.~13593 LNCS, pp.~233--244, 2022.

\bibitem{Karani2018}
N.~Karani, K.~Chaitanya, C.~Baumgartner, and E.~Konukoglu, ``{A lifelong learning approach to brain MR segmentation across scanners and protocols},'' {\em Lecture Notes in Computer Science (including subseries Lecture Notes in Artificial Intelligence and Lecture Notes in Bioinformatics)}, vol.~11070 LNCS, pp.~476--484, 2018.

\bibitem{Huang2019b}
C.~Huang, H.~Han, Q.~Yao, S.~Zhu, and K.~S. Zhou, ``{3D U2-Net: A 3D Universal U-Net for Multi-domain Medical Image Segmentation},'' {\em Proceeding of the International Conference on Medical Image Computing and Computer Assisted Interventions}, vol.~1, pp.~291--299, 2019.

\bibitem{Jay2017}
J.~Y. Zhu, T.~Park, P.~Isola, and A.~A. Efros, ``{Unpaired Image-to-Image Translation Using Cycle-Consistent Adversarial Networks},'' {\em Proceedings of the IEEE International Conference on Computer Vision}, vol.~2017-Octob, pp.~2242--2251, 2017.

\bibitem{Wolterink2017}
J.~M. Wolterink, A.~M. Dinkla, M.~H. Savenije, P.~R. Seevinck, C.~A. van~den Berg, and I.~I{\v{s}}gum, ``{Deep MR to CT synthesis using unpaired data},'' {\em Lecture Notes in Computer Science (including subseries Lecture Notes in Artificial Intelligence and Lecture Notes in Bioinformatics)}, vol.~10557 LNCS, pp.~14--23, 2017.

\bibitem{Armanious2019}
K.~Armanious, C.~Jiang, S.~Abdulatif, T.~K{\"{u}}stner, S.~Gatidis, and B.~Yang, ``{Unsupervised medical image translation using Cycle-MeDGAN},'' {\em European Signal Processing Conference}, vol.~2019-Septe, mar 2019.

\bibitem{Saxena2021}
D.~Saxena and J.~Cao, ``{Generative Adversarial Networks (GANs)},'' {\em ACM Computing Surveys}, vol.~54, no.~3, 2021.

\bibitem{Ho2020}
J.~Ho, A.~Jain, and P.~Abbeel, ``{Denoising diffusion probabilistic models},'' {\em Advances in Neural Information Processing Systems}, vol.~2020-Decem, no.~NeurIPS 2020, pp.~1--25, 2020.

\bibitem{Cottier2024}
B.~Cottier, R.~Rahman, L.~Fattorini, N.~Maslej, and D.~Owen, ``{The rising costs of training frontier AI models},'' {\em arXiv preprint arXiv:2405.21015}, vol.~2016, pp.~1--20, 2024.

\bibitem{Song2021}
J.~Song, C.~Meng, and S.~Ermon, ``{Denoising Diffusion Implicit Models},'' {\em ICLR 2021 - 9th International Conference on Learning Representations}, pp.~1--22, 2021.

\bibitem{Jabri2023}
A.~Jabri, D.~J. Fleet, and T.~Chen, ``{Scalable Adaptive Computation for Iterative Generation},'' {\em Proceedings of Machine Learning Research}, vol.~202, pp.~14569--14589, 2023.

\bibitem{Karras2024}
T.~Karras, M.~Aittala, T.~Kynk{\"{a}}{\"{a}}nniemi, J.~Lehtinen, T.~Aila, and S.~Laine, ``{Guiding a Diffusion Model with a Bad Version of Itself},'' 2024.

\bibitem{Rombach2022}
R.~Rombach, A.~Blattmann, D.~Lorenz, P.~Esser, and B.~Ommer, ``{High-Resolution Image Synthesis with Latent Diffusion Models},'' {\em Proceedings of the IEEE Computer Society Conference on Computer Vision and Pattern Recognition}, vol.~2022-June, pp.~10674--10685, 2022.

\bibitem{Tobin2017}
J.~Tobin, R.~Fong, A.~Ray, J.~Schneider, W.~Zaremba, and P.~Abbeel, ``{Domain randomization for transferring deep neural networks from simulation to the real world},'' {\em IEEE International Conference on Intelligent Robots and Systems}, vol.~2017-Septe, pp.~23--30, 2017.

\bibitem{Bengio2011}
Y.~Bengio, F.~Bastien, A.~Bergeron, N.~Boulanger-Lewandowski, T.~Breuel, Y.~Chherawala, M.~Cisse, M.~C{\^{o}}t{\'{e}}, D.~Erhan, J.~Eustache, X.~Glorot, X.~Muller, S.~P. Lebeuf, R.~Pascanu, S.~Rifai, F.~Savard, and G.~Sicard, ``{Deep learners benefit more from out-of-distribution examples},'' {\em Journal of Machine Learning Research}, vol.~15, pp.~164--172, 2011.

\bibitem{Billot2023}
B.~Billot, D.~N. Greve, O.~Puonti, A.~Thielscher, K.~{Van Leemput}, B.~Fischl, A.~V. Dalca, and J.~E. Iglesias, ``{SynthSeg: Segmentation of brain MRI scans of any contrast and resolution without retraining},'' {\em Medical Image Analysis}, vol.~86, no.~February 2022, 2023.

\bibitem{Tremblay2018}
J.~Tremblay, A.~Prakash, D.~Acuna, M.~Brophy, V.~Jampani, C.~Anil, T.~To, E.~Cameracci, S.~Boochoon, and S.~Birchfield, ``{Training deep networks with synthetic data: Bridging the reality gap by domain randomization},'' {\em IEEE Computer Society Conference on Computer Vision and Pattern Recognition Workshops}, vol.~2018-June, pp.~1082--1090, 2018.

\bibitem{Stojanovski2023}
D.~Stojanovski, U.~Hermida, P.~Lamata, A.~Beqiri, and A.~Gomez, ``{Echo from Noise: Synthetic Ultrasound Image Generation Using Diffusion Models for Real Image Segmentation},'' in {\em Lecture Notes in Computer Science (including subseries Lecture Notes in Artificial Intelligence and Lecture Notes in Bioinformatics)}, vol.~14337 LNCS, pp.~34--43, 2023.

\bibitem{Karras2022}
T.~Karras, M.~Aittala, T.~Aila, and S.~Laine, ``{Elucidating the Design Space of Diffusion-Based Generative Models},'' {\em Advances in Neural Information Processing Systems}, vol.~35, no.~NeurIPS, 2022.

\bibitem{Neal1998}
R.~M. Neal and G.~E. Hinton, ``{A View of the Em Algorithm that Justifies Incremental, Sparse, and other Variants},'' in {\em Learning in Graphical Models}, pp.~355--368, Dordrecht: Springer Netherlands, 1998.

\bibitem{Kingma2014}
D.~P. Kingma and M.~Welling, ``{Auto-encoding variational bayes},'' {\em 2nd International Conference on Learning Representations, ICLR 2014 - Conference Track Proceedings}, no.~Ml, pp.~1--14, 2014.

\bibitem{Deng2011}
Y.~Deng, Y.~Wang, and Y.~Shen, ``{Speckle reduction of ultrasound images based on Rayleigh-trimmed anisotropic diffusion filter},'' {\em Pattern Recognition Letters}, vol.~32, no.~13, pp.~1516--1525, 2011.

\bibitem{Zhang2018}
R.~Zhang, P.~Isola, A.~A. Efros, E.~Shechtman, and O.~Wang, ``{The Unreasonable Effectiveness of Deep Features as a Perceptual Metric},'' no.~1.

\bibitem{Hyvarinen2009}
A.~Hyv{\"{a}}rinen, J.~Hurri, and P.~O. Hoyer, ``{Estimation of Non-normalized Statistical Models},'' pp.~419--426, 2009.

\bibitem{Suli2003}
E.~S{\"{u}}li and D.~F. Mayers, {\em {An Introduction to Numerical Analysis}}.
\newblock Cambridge University Press, aug 2003.

\bibitem{Park2019}
T.~Park, M.~Y. Liu, T.~C. Wang, and J.~Y. Zhu, ``{Semantic image synthesis with spatially-adaptive normalization},'' {\em Proceedings of the IEEE Computer Society Conference on Computer Vision and Pattern Recognition}, vol.~2019-June, pp.~2332--2341, 2019.

\bibitem{Leclerc2019}
S.~Leclerc, E.~Smistad, J.~Pedrosa, A.~Ostvik, F.~Cervenansky, F.~Espinosa, T.~Espeland, E.~A.~R. Berg, P.~M. Jodoin, T.~Grenier, C.~Lartizien, J.~Dhooge, L.~Lovstakken, and O.~Bernard, ``{Deep Learning for Segmentation Using an Open Large-Scale Dataset in 2D Echocardiography},'' {\em IEEE transactions on medical imaging}, vol.~38, no.~9, pp.~2198--2210, 2019.

\bibitem{Kingma2015b}
D.~P. Kingma and J.~L. Ba, ``{Adam: A method for stochastic optimization},'' {\em 3rd International Conference on Learning Representations, ICLR 2015 - Conference Track Proceedings}, pp.~1--15, 2015.

\bibitem{Pinaya2023}
W.~H.~L. Pinaya, M.~S. Graham, E.~Kerfoot, P.-D. Tudosiu, J.~Dafflon, V.~Fernandez, P.~Sanchez, J.~Wolleb, P.~F. da~Costa, A.~Patel, H.~Chung, C.~Zhao, W.~Peng, Z.~Liu, X.~Mei, O.~Lucena, J.~C. Ye, S.~A. Tsaftaris, P.~Dogra, A.~Feng, M.~Modat, P.~Nachev, S.~Ourselin, and M.~J. Cardoso, ``{Generative AI for Medical Imaging: extending the MONAI Framework},'' pp.~1--23, 2023.

\bibitem{Isensee2021}
F.~Isensee, P.~F. Jaeger, S.~A. Kohl, J.~Petersen, and K.~H. Maier-Hein, ``{nnU-Net: a self-configuring method for deep learning-based biomedical image segmentation},'' {\em Nature Methods}, vol.~18, no.~2, pp.~203--211, 2021.

\bibitem{Tan2021}
M.~Tan and Q.~V. Le, ``{EfficientNetV2: Smaller Models and Faster Training},'' {\em Proceedings of Machine Learning Research}, vol.~139, pp.~10096--10106, 2021.

\bibitem{Dice1945}
L.~R. Dice, ``{Measures of the Amount of Ecologic Association Between Species},'' {\em Ecology}, vol.~26, pp.~297--302, jul 1945.

\bibitem{Huttenlocher1992}
D.~P. Huttenlocher, W.~J. Rucklidge, and G.~A. Klanderman, ``{Comparing images using the Hausdorff distance under translation},'' {\em Proceedings of the IEEE Computer Society Conference on Computer Vision and Pattern Recognition}, vol.~1992-June, no.~9, pp.~654--656, 1992.

\bibitem{Paszke2019a}
A.~Paszke, S.~Gross, F.~Massa, A.~Lerer, J.~Bradbury, G.~Chanan, T.~Killeen, Z.~Lin, N.~Gimelshein, L.~Antiga, A.~Desmaison, A.~K{\"{o}}pf, E.~Yang, Z.~DeVito, M.~Raison, A.~Tejani, S.~Chilamkurthy, B.~Steiner, L.~Fang, J.~Bai, and S.~Chintala, ``{PyTorch: An imperative style, high-performance deep learning library},'' tech. rep., 2019.

\end{thebibliography}
\bibliographystyle{ieeetr}

\end{document}